\documentclass[preprint,aps,pra,showpacs,floatfix]{revtex4-1}
\usepackage{graphicx}
\usepackage{caption}
\usepackage{color}
\usepackage{amsmath}
\usepackage{amssymb}
\usepackage{amsfonts}
\usepackage{bm}
\usepackage{hyperref}
\usepackage{url}
\usepackage[english]{babel}
\usepackage[left=2cm,right=2cm,top=2cm,bottom=2cm]{geometry}
\graphicspath{{./figures/}}
\DeclareMathOperator{\arccot}{arccot}

\newcommand{\bfp}{{\bm {p}}}

\newcommand{\bfb}{{\bm {b}}}
\newcommand{\bfe}{{\bm {\eta}}}

\newcommand{\hp}{\hat{\bfp}}

\newcommand{\la}{\langle}
\newcommand{\ra}{\rangle}
\newcommand{\til}[1]{\widetilde {#1}}

\newcommand{\eps}{\ensuremath{\varepsilon}}
\newcommand{\kap}{\ensuremath{\kappa}}

\newcommand{\degree}{\ensuremath{^\circ}}
\newcommand{\vphi}{\ensuremath{\varphi}}

\newcommand{\pbar}{\ensuremath{\bar{p}}}
\renewcommand{\vec}[1]{{\mbox{\boldmath$#1$}}}
\begin{document}
\thispagestyle{empty}
\title{
Relativistic calculations of differential ionization cross sections:
Application to antiproton-hydrogen collisions}
\author{A.~I.~Bondarev$^{1,2}$, Y.~S.~Kozhedub$^{1}$, I.~I.~Tupitsyn$^{1}$, V.~M.~Shabaev$^{1}$, and G.~Plunien$^{3}$}
\affiliation{
$^1$ Department of Physics, St. Petersburg State University,
7/9 Universitetskaya Emb., 199034 St. Petersburg, Russia \\
$^2$ NRC ``Kurchatov Institute'' - ITEP, 
Bolshaya Cheremushkinskaya 25, 117218 Moscow, Russia \\
$^3$ Institut f\"ur Theoretische Physik, Technische Universit\"at Dresden,
Mommsenstra{\ss}e 13, D-01062 Dresden, Germany \\
}
\begin{abstract}
A new relativistic method based on the Dirac equation for calculating fully differential cross sections for ionization in ion-atom collisions is developed. The method is applied to ionization of the atomic hydrogen by antiproton impact, as a non-relativistic benchmark.  The fully differential, as well as various doubly and singly differential cross sections for ionization are presented. The role of the interaction between the projectile and the target nucleus is discussed. Several discrepancies in available theoretical predictions are resolved. The relativistic effects are studied for ionization of hydrogenlike xenon ion under the impact of carbon nuclei.
\end{abstract}

\maketitle
\section{Introduction}
Ionization in ion-atom collisions is of fundamental importance for atomic physics. Within the modern ``Reaction Microscope'' techniques~\cite{doe_00,ull_03}, it is possible to experimentally investigate ionization dynamics at the differential level. The measured fully differential cross sections (FDCS) for ionization being a stringent test of theory stimulate theoretical studies of collisions of ions with atoms and molecules.
\\
Up to date the FDCS for ionization have been successfully measured in collisions involving light targets such as helium~\cite{sch_03,fos_04,gas_16}, lithium~\cite{fis_12,hub_13}, and molecular hydrogen~\cite{cho_11,has_16}. For these targets, non-relativistic theoretical treatment 
is sufficient. However, for heavy targets relativistic effects cannot be neglected, and appropriate target description is required. We note that the relativistic effects induced by fast nuclear motion were investigated for FDCS in the 100~MeV/u C$^{6+}$-He collision in Refs.~\cite{mcg_10b,kou_12}. The Facility for Antiproton and Ion Research (FAIR) being constructed in Darmstadt (Germany)~\cite{fair} will be able to provide heavy ions up to bare uranium and has an extensive scientific program aimed at the research of heavy-ion collision dynamics.
\\
At the same time, we would like to point out promising plans for antiproton research at the FAIR~\cite{wid_15} and the European Organization for Nuclear Research (CERN)~\cite{mau_14}. Experimental and theoretical studies of ionization processes in collisions of antiprotons with atoms and molecules have been recently reviewed~\cite{kir_11}. Despite of the lack of the experimental FDCS, the essentially non-relativistic antiproton-hydrogen collision serves as an ideal benchmark to various theories. This is due to the absence of the charged-transfer channel, in contrast to the collisions with protons, and absence of the electron correlations, in contrast to the electron-impact or multi-electron-target ionization. 
\\
The perturbative calculations of FDCS for ionization in antiproton-hydrogen collision have been performed in Refs.~\cite{ber_93,jon_02,voi_03}. Recently these FDCS have also been studied by several non-perturbative methods~\cite{mcg_09,mcg_10,abd_11,cia_13,abd_16}. Firstly, McGovern {\it et al.}~\cite{mcg_09,mcg_10} developed a method for extracting the FDCS from an impact-parameter treatment of the collision within a coupled pseudostate (CP) formalism. Later, Abdurakhmanov {\it et al.}~\cite{abd_11} worked out the fully quantal time-independent convergent-close-coupling (QM-CCC) approach to differential ionization studies in ion-atom collisions. Recently, Ciappina {\it et al.}~\cite{cia_13} applied the time-dependent close-coupling (TDCC) technique to investigate the role of the nucleus-nucleus interaction in the FDCS. Afterwards, Abdurakhmanov {\it et al.}~\cite{abd_16} used the semiclassical wave-packet convergent-close-coupling (WP-CCC) method to examine the FDCS. We also would like to mention the recent paper by Sarkadi and Guly\'{a}s~\cite{sar_14}, where the FDCS were investigated using the classical-trajectory Monte Carlo method. 
\\
In this contribution, we present a relativistic single-center semiclassical coupled-channel approach based on the Dirac equation to calculation FDCS for ionization in ion-atom collisions. The basis of target pseudostates is used for the scattering wave function expansion. These pseudostates representing bound states as well as discretized positive- and negative-energy Dirac continua are obtained by diagonalization of the target Hamiltonian utilizing {\it B}-splines. {\it B}-splines were introduced in atomic physics calculations in 1970s and are broadly used in various problems~(see, e.g., reviews~\cite{sap_96,bac_01}). In particular, the {\it B}-spline expansion has already been applied to calculate the total ionization cross sections in antiproton-hydrogen collisions in Refs.~\cite{azu_01,sah_04}. 
We report the results of the developed method application to the problem of antiproton-impact ionization of atomic hydrogen, where noticeable disagreements in available theoretical predictions exist. We also report the total ionization probabilities of hydrogenlike xenon ion under the impact of carbon nuclei to demonstrate importance of the relativistic effects. 
\\
\indent The paper is organized as follows. In Sec.~\ref{s:theory} the relativistic method is described. Details of the calculations are given in Sec.~\ref{ss:details}. The results for the \pbar-H and C$^{6+}$-Xe$^{53+}$ collisions are presented in Secs.~\ref{ss:pbar-H} and~\ref{ss:C-Xe}, respectively. In Sec.~\ref{s:conclusion} we give the conclusions. Atomic units (a.u.) $\hbar = e = m_{e} = 1$ are used throughout the paper unless otherwise stated. 
\section{Theory} \label{s:theory}
\subsection{Time-dependent Dirac equation in a finite basis set} \label{ss:TDDE}
We consider the collision of a one-active-electron target with a bare projectile. Within the semiclassical approximation, we treat the nuclei as sources of an external time-dependent potential. 
Thus the many-particle problem is reduced to the motion of the relativistic electron in a two-center time-dependent potential. The electron dynamics is described by the time-dependent Dirac equation, 
\begin{equation} \label{eq:dirac}
i\frac{\partial \Psi(\vec{r},t,\vec{R})}{\partial t} = H(t)\Psi(\vec{r},t,\vec{R}),
\end{equation}
where the total Hamiltonian of the system is the sum of the free relativistic Dirac Hamiltonian and the interactions between the active electron with the target atom and the projectile, and is given by
\begin{gather} \label{eq:hamiltonian}
H(t) =  H_0+V_{\rm{P}}(t), \\
H_0 = c(\vec{\alpha} \cdot \vec{p})+(\beta-1) c^2+V_{\rm T},
\end{gather}
where $\vec{\alpha}$ and $\beta$ are the Dirac matrices.
Let us assume that the target is located at the origin, while the projectile moves along a straight-line trajectory $\vec{R} = \vec{b}+\vec{v}t$ with the constant velocity $\vec{v}$ and at the impact parameter $\vec{b}$, so that $\vec{b}\cdot\vec{v} = 0$. Then the total two-center potential $V(\vec{r},t)$ is written as 
\begin{equation} \label{eq:potential}
V(\vec{r},t) = V_{\rm T}(r)+V_{\rm P}(|\vec{r}-\vec{R}(t)|).
\end{equation} 
We note that the potential $V(\vec{r},t)$ does not include the interaction between the target nucleus and the projectile. This nucleus-nucleus (NN) interaction does not affect cross sections, which are not differential in the scattered projectile variables. For cross sections, which are differential in these variables, it can be taken into account by a phase transformation in Eq.~\eqref{eq:dirac}.
The target potential $V_{\rm T}$ consists of the Coulomb potential of the nucleus $V_{\rm nucl}$ and the screening potential of the passive electrons $V_{\rm scr}$,
\begin{equation}
V_{\rm T} = V_{\rm nucl}+V_{\rm scr}.
\end{equation}
The finite nuclear size effects are incorporated in $V_{\rm nucl}$ using an appropriate nuclear charge distribution. The local screening potential of the passive electrons $V_{\rm scr}$ can be obtained using various approximate methods. 
\\
To solve Eq.~\eqref{eq:dirac}, we expand the time-dependent wave function $\Psi(\vec{r},t,\vec{R})$ over a finite basis set,
\begin{equation} \label{eq:basis_exp}
\Psi(\vec{r},t, \vec{R}) =  \sum_{a}C_{a}(t,\vec{b}) e^{-i\eps_{a}t}\vphi_{a}(\vec{r}), 
\end{equation}
where the basis functions $\vphi_a$ are orthonormal and obtained by diagonalization of the stationary atomic Hamiltonian $H_0$ employing {\it B}-splines~\cite{joh_88,bon_13},
\begin{equation} \label{eq:basis_const}
\la \vphi_a | H_0 | \vphi_a \ra = \eps_a, \qquad \la \vphi_a | \vphi_b \ra = \delta_{ab}.
\end{equation}
Since the target potential $V_{\rm T}(r)$ possesses the spherical symmetry, the basis function $\vphi_a(\vec{r})$ may be represented as the bispinor $\vphi_{n_a \kap_a \mu_a}(\vec{r})$ with a given principal quantum number $n_a$, angular momentum-parity quantum number $\kap_a = (-1)^{l_a+j_a+1/2}(j_a+1/2)$, and angular momentum projection on the $z$-axis $\mu_a$,
\begin{equation} 
  \vphi_a(\vec{r}) \equiv \vphi_{n_a \kap_a \mu_a}(\vec{r}) = \frac{1}{r} \begin{pmatrix} 
					G_{n_a \kap_a}(r)\, \chi_{\kap_a \mu_a}(\hat{\vec{r}}) \\
					i\, F_{n_a \kap_a}(r)\, \chi_{-\kap_a \mu_a}(\hat{\vec{r}})
			      \end{pmatrix},
\end{equation} 
where $G_{n_a\kap_a}(r)$ and $F_{n_a\kap_a}(r)$ are the large and small radial components, respectively, and $\chi_{\kap_a\mu_a}(\hat{\vec{r}})$ are the spherical spinors, and $\hat{\vec{r}} = \vec{r}/r$~\cite{ros_61}. In the following, we assume that the $z$-axis is directed along the vector $\vec{v}$.
\\
The basis functions $\vphi_a$ represent bound states, positive-energy, as well as negative-energy Dirac continuum. Moreover, for low-lying bound states they are very close to the exact ones. Their quality and overall number depends on the size of the {\it B}-spline basis set.
We note that due to using the dual-kinetic-balance approach~\cite{sha_04}, the basis set 
is free from the so-called spurious states, which may arise in a finite-basis-set representation of the Dirac equation~\cite{tup_08}.
\\
Substituting Eq.~\eqref{eq:basis_exp} into Eq.~\eqref{eq:dirac}, one derives the set of coupled-channel equations for the expansion coefficients,
\begin{equation}  \label{eq:cc}
i\frac{dC_a(t,\vec{b})}{dt} = \sum_{b} C_{b}(t,\vec{b}) e^{i(\eps_a-\eps_b)t }\la\vphi_a|V_{\rm{P}}|\vphi_b\ra 
\end{equation} 
with the initial conditions corresponding to the initial active electron state~$i$,
\begin{equation} \label{eq:init_cond}
C_a(t \to-\infty,\vec{b}) = \delta_{ai}.
\end{equation}
It should be noted that the atomic-like basis set centered at the target does not allow for the explicit description of charge transfer processes. So the method is reliable, if the charge transfer processes are minor compared to the direct ionization ones. This condition is met for fast projectiles, relatively (compared to the target) light projectiles, and projectiles without electron bound states.
\\
From the properties of the matrix element $V_{ab}(\vec{R}) \equiv \langle\vphi_a|V_{\rm{P}}|\vphi_b\rangle$ under rotation around the $z$-axis, it follows that
\begin{equation}
V_{ab}(\vec{R}) = \til{V}_{ab}(t,b)e^{i(\mu_b-\mu_a)\phi_b},
\end{equation}
where $\phi_b$ is the azimuthal angle of $\vec{b}$.
Then the dependence of the expansion coefficient $C_a(t,\vec{b})$ on $\phi_b$ can also be factorized,
\begin{equation}
C_a(t,\vec{b}) = \til{C}_a(t,b)e^{i(\mu_i-\mu_a)\phi_b},
\end{equation}
where $\til{C}_a(t,b)$ satisfies the system
\begin{equation}  \label{eq:cc'}
i\frac{d\til{C}_a(t,b)}{dt} = \sum_{b} \til{C}_b(t,b) e^{i(\eps_a-\eps_b)t}\til{V}_{ab}(t,b)
\end{equation} 
with the initial conditions 
\begin{equation}
\til{C}_a(t \to -\infty,b) = \delta_{ai}.
\end{equation}
To evaluate the matrix elements, it is convenient to reexpand the potential of the projectile to the target position, where the basis functions are centered. If the finite nuclear size effect for the projectile is neglected, the reexpansion of its Coulomb potential can be done analytically~\cite{var_88}, 
\begin{equation} \label{eq:reexpanding}
-\frac{Z_{\rm P}}{|\vec{r}-\vec{R}|} = -\frac{Z_{\rm P}}{r_>}\sum_{l=0}^{\infty}\biggl(\frac{r_<}{r_>}\biggr)^l \! \sum_{m=-l}^{l}C^{l}_{m}(\hat{\vec{r}})C_{m}^{l*}(\hat{\vec{R}}),
\end{equation}
where $r_<$ and  $r_>$ are the minimum and maximum values of $(r,R)$, respectively, and $C^{l}_{m}$ denotes the spherical tensor, which is related to the spherical harmonic $Y_{lm}$ as
\begin{equation}
C^{l}_{m}(\hat{\vec{r}}) = \sqrt{\frac{4\pi}{2l+1}}Y_{lm}(\hat{\vec{r}}).
\end{equation}
Thus the matrix element $\til{V}_{ab}$ may be represented in the following form: 
\begin{equation} 
\til{V}_{ab}(t,b) \equiv \til{V}_{n_a\kap_a\mu_a n_b\kap_b\mu_b}(t,b) = \sum_{lm} R^{l}_{n_a\kap_a n_b\kap_b}(t,b)\, A^{lm}_{\kap_a \mu_a \kap_b \mu_b}\,C^{l*}_{m}(\arccot vt/b,0),
\end{equation}
where the radial part is given by
\begin{equation} 
R^{l}_{n_a\kap_a n_b\kap_b} = -Z_{\rm P} \int_{0}^{\infty} \! dr \frac{1}{r_>}\left(\frac{r_{<}}{r_{>}}\right)^l \Bigl[G_{n_a\kap_a}(r) G_{n_b\kap_b}(r) + F_{n_a\kap_a}(r) F_{n_b\kap_b}(r)\Bigr],
\end{equation}
and the angular part is the so-called relativistic Gaunt coefficient,
\begin{equation} 
A^{lm}_{\kap_a \mu_a \kap_b \mu_b} = \la \chi_{\kap_a \mu_a}|C^{l}_{m}| \chi_{\kap_b \mu_b} \ra = g^{lm}(j_a \mu_a; \, j_b \mu_b).
\end{equation}
It may be expressed through the $3j$-symbols as
\begin{subequations} \label{eq:gaunt_definition}
\begin{equation} 
g^{lm}(j_a \mu_a; \, j_b \mu_b) =  (-1)^{\frac{1}{2}+\mu_a}\, \sqrt{(2j_a+1)(2j_b+1)}
 \begin{pmatrix}
j_a  & l & j_b \\
\frac{1}{2}  & 0 & -\frac{1}{2} \\
\end{pmatrix} 
\begin{pmatrix}
j_a  & l & j_b \\
-\mu_a  & m & \mu_b
\end{pmatrix},
\end{equation}
where $l_a+l_b+l$ should be even number, otherwise 
\begin{equation} 
g^{lm}(j_a \mu_a; \, j_b \mu_b) = 0. 
\end{equation}
\end{subequations}
The commonly used non-relativistic Gaunt coefficient is proportional to the well-known integral of three spherical harmonics~\cite{var_88}.
\\
We note that here the matrix elements are calculated in the laboratory reference frame.
From the computational point of view, this is not the most efficient way. There are two alternative possibilities. One may calculate them in the local reference frame, where $z$-axis is parallel to the internuclear vector $\vec{R}$ at each time moment. Then one should either rotate these matrix elements from the local to the laboratory reference frame using the Wigner D-functions (see, e.g., Ref.~\cite{tup_10}), or rewrite the time-dependent equation~\eqref{eq:dirac} in this local rotating reference frame. Since the rotating reference frame is non-inertial, an additional term arises in the Hamiltonian~\eqref{eq:hamiltonian} (see, e.g., Ref.~\cite{mal_13}).
\\
We also would like to mention the symmetry properties of the matrix elements, which can be used for their calculation and storage:
\begin{equation}
\til{V}_{n_a\kap_a-\mu_a n_b\kap_b-\mu_b} = (-1)^{(j_b+l_b+\mu_b-j_a-l_a-\mu_a)} \til{V}_{n_a\kap_a\mu_a n_b\kap_b\mu_b}.
\end{equation}
\\
The system of equations~\eqref{eq:cc'} may be rewritten in the matrix form,
\begin{equation} \label{eq:matrix}
i\dfrac{d\vec{\til{C}}}{dt} = M\vec{\til{C}}, \quad  M_{ab} = e^{i(\eps_a-\eps_b)t}\til{V}_{ab},
\end{equation}
where $\vec{\til{C}}$ is the vector incorporating the expansion coefficients $\til{C}_{a}$.
To solve Eq.~\eqref{eq:matrix}, we use the short iterative Lanczos propagator~\cite{par_86,lef_91}.
It is an exponential-type propagator, where the matrix exponential is approximated in the Krylov subspace~\cite{mol_03}. The Lanczos propagation is a standard procedure widely used in various chemical and physical calculations~\cite{fei_08,gol_15}. 
\subsection{Cross sections} \label{ss:TAP}
The total ionization probability is calculated as the following sum over the positive-energy basis states:
\begin{equation}\label{eq:total_p_ion}
P_{\rm ion}(\vec{b}) = P_{\rm ion}(b) = \! \! \sum_a^{(\eps_a > 0)} \! \! |\til{C}_a(t\to\infty,b)|^2.
\end{equation}
An alternative method used in Refs.~\cite{mcg_09,abd_11a}, where the summation runs over all basis states and for each of them the overlap with the positive-energy continuum is taken into account, gives almost the same results in a sufficiently large basis set.
\\
The total ionization cross section follows from
\begin{equation}\label{eq:total_sigma_ion}
\sigma_{\rm ion} = \int \! d\vec{b}\, P_{\rm ion}(\vec{b}) = 2\pi \int_{0}^{\infty} \! db\, b\, P_{\rm ion}(b).
\end{equation}
Using the Stieltjes technique for every symmetry $\kap_a$, we are also able to calculate partial transition probabilities differential in the energy of the electron~\cite{bon_15},
\begin{equation} \label{eq:dp_discr}
\frac{dP^{\kap_a}_{\rm tr}}{d\eps}\biggl(\frac{\eps_{n_a+1}^{\kap_a}+\eps_{n_a}^{\kap_a}}{2},b\biggr) =  \frac{1}{2}\frac{P_{n_a+1}^{\kap_a}(b)+P_{n_a}^{\kap_a}(b)}{\eps_{n_a+1}^{\kap_a}-\eps_{n_a}^{\kap_a}}, \quad P_{n_a}^{\kap_a}(b) = \sum_{\mu_a}|\til{C}_{n_a\kap_a\mu_a}(t\to\infty,b)|^2.
\end{equation}
After interpolation of the partial probabilities on a common energy grid, summation over the symmetries and integration over the impact parameter, one obtains the single differential cross section for the transition,
\begin{equation} \label{eq:dsig_de_discr}
\frac{d\sigma_{\rm tr}}{d\eps} = 2\pi \int_{0}^{\infty} \! db\, b\, \sum_{\kap_a}\frac{dP^{\kap_a}_{\rm tr}(b)}{d\eps}.
\end{equation}
We note that the energies $\eps_a \equiv \eps_{n_a}^{\kap_a}$ are obtained by diagonalization of the stationary atomic Hamiltonian $H_0$ in the finite {\it B}-spline set (see Eq.~\eqref{eq:basis_const}) and can not be chosen arbitrary. Moreover, basis functions $\vphi_a$ with energy $\eps_a$ near the ionization threshold have a similar behavior for positive and negative values of energy $\eps_a$. Thus Eq.~\eqref{eq:dp_discr} can be used for $\eps_a < 0$ as well, giving in this case the excitation probability into an energy interval, in contrast to the differential ionization probability for $\eps_a > 0$.
\\
We proceed with evaluation of the probability of the electron ejection in a given direction. The spherical-wave decomposition of the outgoing continuum electron wave function $\Psi^{(-)}_{\eps \hp \mu_s}(\vec{r})$ with a given asymptotic momentum $\vec{p}$ and spin projection at the $z$-axis $\mu_s$ is~\cite{eic_07}
\begin{equation}\label{eq:sph_decomp}
\Psi^{(-)}_{\eps \hp \mu_s}(\vec{r}) = \sum_{\kap\mu m} i^l\, e^{-i\Delta_\kap}\,C^{j\mu}_{lm,\frac{1}{2}\mu_s}\,Y^{*}_{lm}(\hp)\,\psi_{\eps\kap\mu}(\vec{r}),
\end{equation}
where $C^{j\mu}_{lm,\frac{1}{2}\mu_s}$ is the Clebsch-Gordan coefficient, $\psi_{\eps\kap\mu}$ is the Dirac partial wave, and $\Delta_\kap$ is the difference between the asymptotic large-distance phase of the Dirac-Coulomb solution and the free Dirac solution~\cite{ros_61}. The Dirac partial wave $\psi_{\eps\kap\mu}$ with a given energy $\eps$, angular momentum-parity quantum number $\kap$, and angular momentum projection $\mu$ is represented by
\begin{equation} \label{eq:salvat_wf}
\psi_{\eps\kap\mu}(\vec{r}) = \frac{1}{r}\begin{pmatrix}
				      G_{\eps\kap}(r)\, \chi_{ \kap\mu}(\hat{\vec{r}}) \\
				  i\, F_{\eps\kap}(r)\, \chi_{-\kap\mu}(\hat{\vec{r}})
		  		\end{pmatrix},
\end{equation}
and normalized on the energy scale,
\begin{equation}
\la \psi_{\eps\kap\mu} | \psi_{\eps'\kap\mu} \ra = \delta(\eps-\eps').
\end{equation}
The radial components $G_{\eps\kap}$ and $F_{\eps\kap}$ of the wave function $\psi_{\eps\kap\mu}$ and the phase shift $\Delta_\kap$  are obtained using the RADIAL package~\cite{sal_95}. In contrast to the energies $\eps_a$ used in Eq.~\eqref{eq:dp_discr}, $\eps$ may be chosen arbitrary.
 Note that since we quantize the spin of the ejected electron in the direction of the $z$-axis, the summation over $\mu_s = \pm 1/2$ is required in final expressions for observables. Alternatively, one may quantize the spin of the ejected electron in the direction of its propagation. Then the components with different projections (helicities) can be obtained and, in principle, measured. 
\\
The transition amplitude $T^{\mu_s}(\eps, \theta_e, \phi_e, b, \phi_b)$ is obtained projecting the wave function $\Psi(\vec{r},t,\vec{R})$ on the wave function $\Psi^{(-)}_{\eps \hp \mu_s}(\vec{r})$ at the asymptotic time,
\begin{equation} \label{eq:projection}
T^{\mu_s}(\eps, \theta_e, \phi_e, b, \phi_b) = \la \Psi^{(-)}_{\eps \hp \mu_s}\, e^{-i\eps t} | \Psi \ra, \quad t \to \infty,
\end{equation}
where the angles $\theta_e$ and $\phi_e$ correspond to the direction $\hp$ of the ejected electron. Projecting $\Psi^{(-)}_{\eps \hp \mu_s}$ onto the basis states $\vphi_a$ and using their orthonormality (see Ref.~\cite{bon_15} for details) we come to the following expression for the transition amplitude:
\begin{equation} \label{eq:trans_ampl}
T^{\mu_s}(\eps, \theta_e, \phi_e, b, \phi_b) = \sum_{\kap}(-i)^l e^{i\Delta_{\kap}} \sum_{\mu m} C^{j\mu}_{lm\frac{1}{2}\mu_s} Y_{lm}(\theta_e,\phi_e) e^{i(\mu_i-\mu)\phi_b}\sum_{n} I^{\kap}_{\eps n} \til{C}_{n\kap\mu}(t\to\infty,b),
\end{equation}
where $\mu_i$ is the angular momentum projection of the initial state and the radial overlapping integral $I_{\eps n}^\kap$ is given by
\begin{equation}
I^{\kap}_{\eps n} = \int_{0}^{\infty} \! dr  \,[G_{\eps\kap}(r)G_{n\kap}(r)+F_{\eps\kap}(r)F_{n\kap}(r)].
\end{equation}
Then the fully differential ionization probability as a function of the impact parameter $\vec{b}$, the electron ejection energy $\eps$, and the electron ejection angles $\theta_e$ and $\phi_e$ is given by
\begin{equation} \label{eq:d3P_b}
\frac{d^3P(\vec{b})}{d\eps\, d(\cos\theta_e)\, d\phi_e} = \sum_{\mu_s = \pm\frac{1}{2}}|T^{\mu_s}(\eps, \theta_e, \phi_e, b, \phi_b)|^2.
\end{equation}
We note that in the non-relativistic limit, the electron spin projection at any axis is conserved and, as a result, one term in Eq.~\eqref{eq:d3P_b} vanishes.
\\
For comparison with an experiment, it is usually more convenient to express the differential probabilities in terms of the transverse  (perpendicular to $\vec{v}$)  component $\vec{\eta}$ of the projectile momentum transfer $\vec{q}$ rather than the impact parameter $\vec{b}$. The projectile momentum transfer is the difference between the initial ($\vec{k_i}$) and final ($\vec{k_f}$) projectile momenta $\vec{q} = \vec{k_i}-\vec{k_f}$.  
\\
Transition amplitudes in the $b$- and $\eta$-representations are related by a two-dimensional Fourier transform~\cite{mcd_70,eic_95},
\begin{equation} \label{eq:FT_gen}
T^{\mu_s}(\eps,\theta_e,\phi_e,\eta,\phi_\eta) = \frac{1}{2\pi}\int \! d\vec{b}\, e^{i\bfe\cdot\bfb}\, e^{i\delta(b)}\,  T^{\mu_s}(\eps,\theta_e,\phi_e,b,\phi_b),
\end{equation}
where $\delta(b)$ is the additional phase due to the NN interaction omitted in Eq.~\eqref{eq:potential}. This phase depends on the explicit form of the NN interaction, which may include the Coulomb interaction between the projectile and the target nucleus, the projectile and the passive target electrons, as well as polarization effects. In the simple approximation, where the presence of the passive target electrons is accounted for by changing the target charge $Z_{\rm T}$ to some screened value $Z_{\rm eff}$,
\begin{equation} \label{eq:NNeff}
V_{NN}(R) = \frac{Z_{\rm eff}Z_{\rm P}}{R}.
\end{equation}
In this case, the phase factor $\delta(b)$ reads as
\begin{equation} \label{eq:NN_phase}
\delta(b) = \frac{2Z_{\rm{eff}}Z_{\rm{P}}}{v}\ln{vb}.
\end{equation}
Some useful remarks on the derivation and applicability of this expression can be found in Ref.~\cite{wal_12}. Moreover, in the present calculations, we explicitly checked that inclusion of the NN interaction~\eqref{eq:NNeff} directly in Eq.~\eqref{eq:potential} or as the phase factor~\eqref{eq:NN_phase} in Eq.~\eqref{eq:FT_gen} gives indistinguishable results.
\\
Using the Jacobi-Anger expansion~\cite{abr_72}, we express the Fourier transform of the amplitude $T^{\mu_s}(\eps,\theta_e,\phi_e,b,\phi_b)$ as
\begin{equation} \label{eq:FT}
T^{\mu_s}(\eps,\theta_e,\phi_e,\eta,\phi_\eta) = \frac{1}{2\pi}\int_{0}^{2\pi} \! d\phi_b \int_{0}^{\infty} \! b\, db \sum_n i^n \,  e^{in(\phi_b-\phi_\eta)}\,  J_n(\eta b)\, e^{i\delta(b)}\,  T^{\mu_s}(\eps,\theta_e,\phi_e,b,\phi_b).
\end{equation}
Here $J_n(\eta b)$ is the $n$-th order Bessel function of the first kind and $\phi_\eta$ is the azimuthal angle of the transverse component of the momentum transfer $\vec{\eta}$.
The integration over $\phi_b$ gives 
\begin{equation} \label{eq:FT_simp}
T^{\mu_s}(\eps,\theta_e,\phi_e,\eta,\phi_\eta) = \sum_{\kap}(-i)^l e^{i\Delta_{\kap}} \sum_{\mu m} C^{j\mu}_{lm\frac{1}{2}\mu_s} Y_{lm}(\theta_e,\phi_e)\, i^{(\mu-\mu_i)}\, e^{i(\mu_i-\mu)\phi_\eta}\sum_{n} I^{\kap}_{\eps n} B^{\mu-\mu_i}_{n\kap\mu}(\eta),
\end{equation}
where
\begin{equation} \label{eq:coef_four_trans}
B^m_{n\kap\mu}(\eta) = \int_{0}^{\infty} \! b\, db\, J_m(\eta b)\, e^{i\delta(b)}\, \til{C}_{n\kap\mu}(t\to\infty,b).
\end{equation}
Then the fully differential ionization probability as a function of the transverse component of the momentum transfer $\vec{\eta}$, the electron ejection energy $\eps$, and the electron ejection angles $\theta_e$ and $\phi_e$  is calculated as
\begin{equation} \label{eq:fdcs-eta}
\frac{d^3P(\vec{\eta})}{d\eps\, d(\cos\theta_e)\, d\phi_e} = \sum_{\mu_s = \pm\frac{1}{2}}|T^{\mu_s}(\eps,\theta_e,\phi_e,\eta,\phi_\eta)|^2.
\end{equation}
The (fully) triply differential cross section (TDCS) may be expressed as
\begin{equation} \label{eq:fdcs-eta-renorm}
\frac{d^3\sigma}{d\eps\, d\Omega_e\, d\Omega_{\rm P}} = k_i k_f \frac{d^3P(\vec{\eta})}{d\eps\, d(\cos\theta_e)\, d\phi_e}.
\end{equation}
This is the cross section for the electron being ejected with the energy in the range from $\eps$ to $\eps+d\eps$ into the solid angle $d\Omega_e$, while the projectile is scattered into the solid angle $d\Omega_{\rm P}$. It depends on the reference frame through the initial and final projectile momenta, since the solid angle $d\Omega_{\rm{P}}$ is different in the laboratory and center of mass reference frames. \\
Integrating the TDCS over corresponding variables, one can obtain various doubly differential cross sections (DDCS), singly differential cross sections (SDCS), and, finally, the total ionization cross section.
From the sets of DDCS and SDCS, we focus here only on those, in which significant disagreements with the previously published results have been found. These are the DDCS $\frac{d^2\sigma}{d\eps \, d\eta}$ and SDCS $\frac{d\sigma}{d\eps}$. The former is defined by 
\begin{equation} \label{eq:d2sigdedeta_def}
\frac{d^2\sigma}{d\eps\, d\eta} = \frac{\eta}{k_i k_f} \int_0^{2\pi} \! \frac{d^2\sigma}{d\eps\, d\Omega_{\rm P}} d\phi_{\rm P},
\end{equation}
where
\begin{equation}
 \frac{d^2\sigma}{d\eps\, d\Omega_{\rm P}} =  \int \! \frac{d^3\sigma}{d\eps\, d\Omega_e\, d\Omega_{\rm P}} d\Omega_e
\end{equation}
and $\phi_{\rm P}$ is the azimuthal angle of the scattered projectile. In our approach, it can be calculated as
\begin{equation} \label{eq:d2sigdedeta_cal}
\frac{d^2\sigma}{d\eps\, d\eta} = \eta \int_0^{2\pi} \! d\phi_\eta\, \int_{-1}^{1} \! d(\cos\theta_e) \int_0^{2\pi} \! d\phi_e \frac{d^3P(\vec{\eta})}{d\eps\, d(\cos\theta_e)\, d\phi_e}.
\end{equation}
The latter is defined as
\begin{equation}
\frac{d\sigma}{d\eps} = \int \! d\Omega_e \int \! d\Omega_{\rm P}\, \frac{d^3\sigma}{d\eps\, d\Omega_e\, d\Omega_{\rm P}}
\end{equation}
and can be easier calculated in the $b$- rather than in $\eta$-representation as
\begin{equation} \label{eq:dsig_de}
\frac{d\sigma}{d\eps} =  \int_0^\infty \! b\, db\, \int_0^{2\pi} \! d\phi_b\, \int_{-1}^1 \! d(\cos\theta_e)\, \int_0^{2\pi} \! d\phi_e\, \frac{d^3P(\vec{b})}{d\eps\, d(\cos\theta_e)\, d\phi_e}.
\end{equation}
We note that due to the symmetry, the fully differential ionization probability $\frac{d^3P(\vec{b})}{d\eps\, d(\cos\theta_e)\, d\phi_e}$ depends on $\phi_b$ and $\phi_e$ only through their difference, and the integration in Eq.~\eqref{eq:dsig_de} can be simplified. The similar arguments for the $\frac{d^3P(\vec{\eta})}{d\eps\, d(\cos\theta_e)\, d\phi_e}$  allow us to simplify the integration in Eq.~\eqref{eq:d2sigdedeta_cal}.
The SDCS calculated by means of Eqs.~\eqref{eq:dsig_de} and~\eqref{eq:dsig_de_discr} for positive energies~$\eps_a$ should be the same. This criterion can serve for checking of the calculations involving the wave function of the ejected electron defined by Eq.~\eqref{eq:sph_decomp}.
\\
A useful check for the convergence over the basis set size is to obtain the first-order perturbative solution of the coupled-channel equations~\eqref{eq:cc}:
\begin{equation} \label{eq:B1_coef}
C^{\rm B1}_{a}(t,\vec{b}) = \delta_{ai}-i\int_{-\infty}^{t} \! dt' e^{i(\eps_a-\eps_i)t'} \la \vphi_a | V_{\rm P} | \vphi_i \ra.
\end{equation}
Cross sections calculated using this perturbative solution should then be compared with the corresponding cross sections in the first Born approximation (FBA). We note that in the FBA, the NN interaction does not contribute to the cross sections due to orthogonality of the wave functions in Eq.~\eqref{eq:B1_coef}.
\section{Results}  
\subsection{Details of calculations}  \label{ss:details}
We used the theory described above to calculate cross sections for ionization in the antiproton-hydrogen collision. In the present calculation for the antiproton-hydrogen collision, we did not include in the expansion~\eqref{eq:basis_exp} negative-energy continuum states, which result from the target Hamiltonian diagonalization. Furthermore, we omitted high-energy states with $\eps_k > 10$~a.u. With these restrictions, the basis set consisted of $45$ radial functions for each angular symmetry. The  states with the angular momentum-parity quantum number $\kap = \pm1,\ldots,\pm8,-9$, which corresponds to  $l = 0,\ldots,8$, were included in the basis set. The coupled-channel equations~\eqref{eq:cc'} were solved from $z_{\rm min} = -60$~a.u. to $z_{\rm max} = 60$~a.u., where $z=vt$ is the $z$-component of the projectile position.
\subsection{Antiproton-impact ionization of atomic hydrogen} \label{ss:pbar-H}
\begin{figure}[htb]
\includegraphics[width=0.98\linewidth]{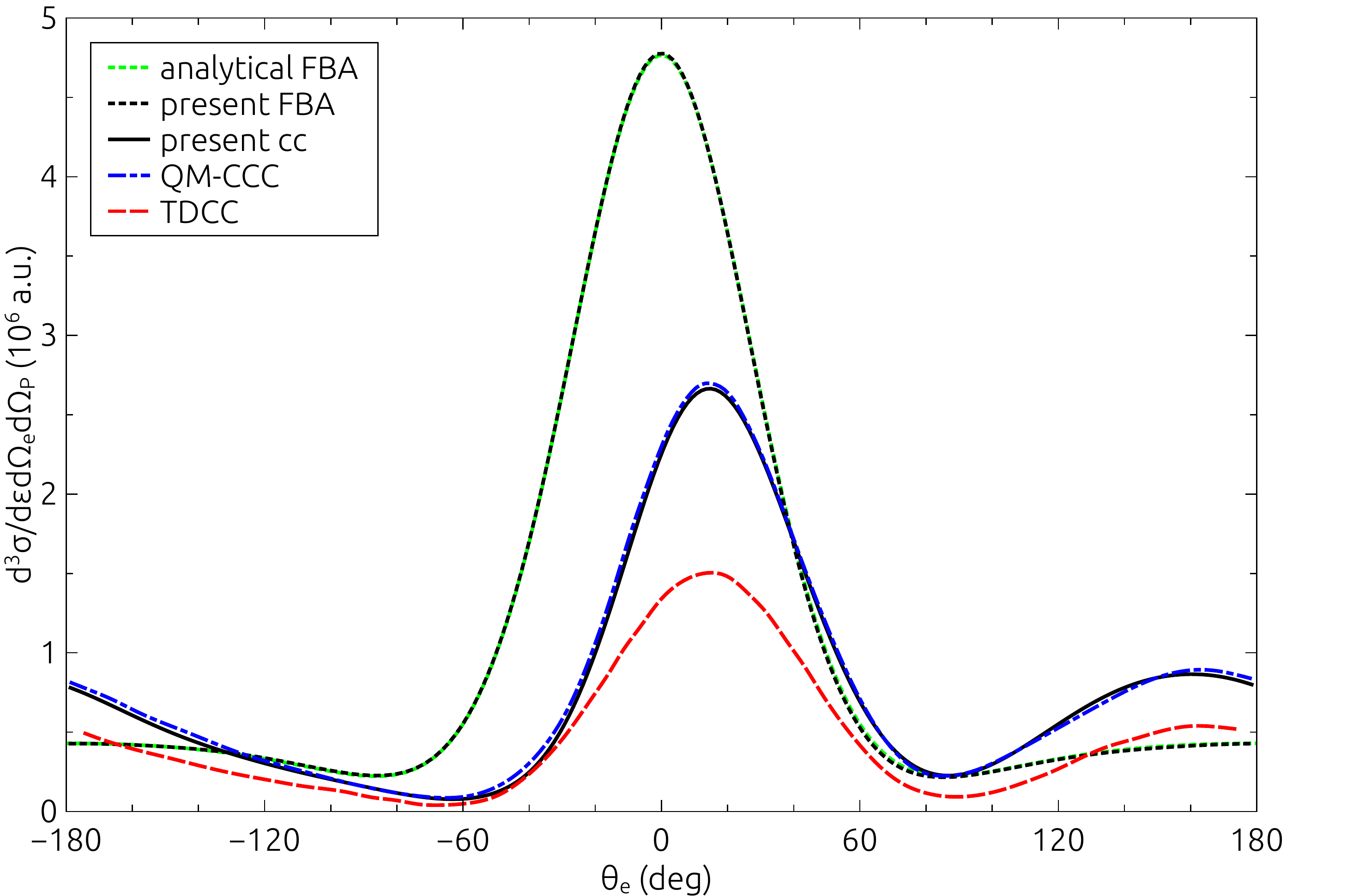}
\caption{TDCS for antiproton-impact ionization of hydrogen at $200$~keV in the scattering plane. The scattering angle of the projectile is $0.2$~mrad and the ejected electron energy is $4$~eV. The results of the QM-CCC and TDCC approaches are from Refs.~\cite{abd_11} and~\cite{cia_13}, respectively.}
\label{fig:G1_200keV_04eV_02mrad}
\end{figure}

\begin{figure}[htb]
\includegraphics[width=0.98\linewidth]{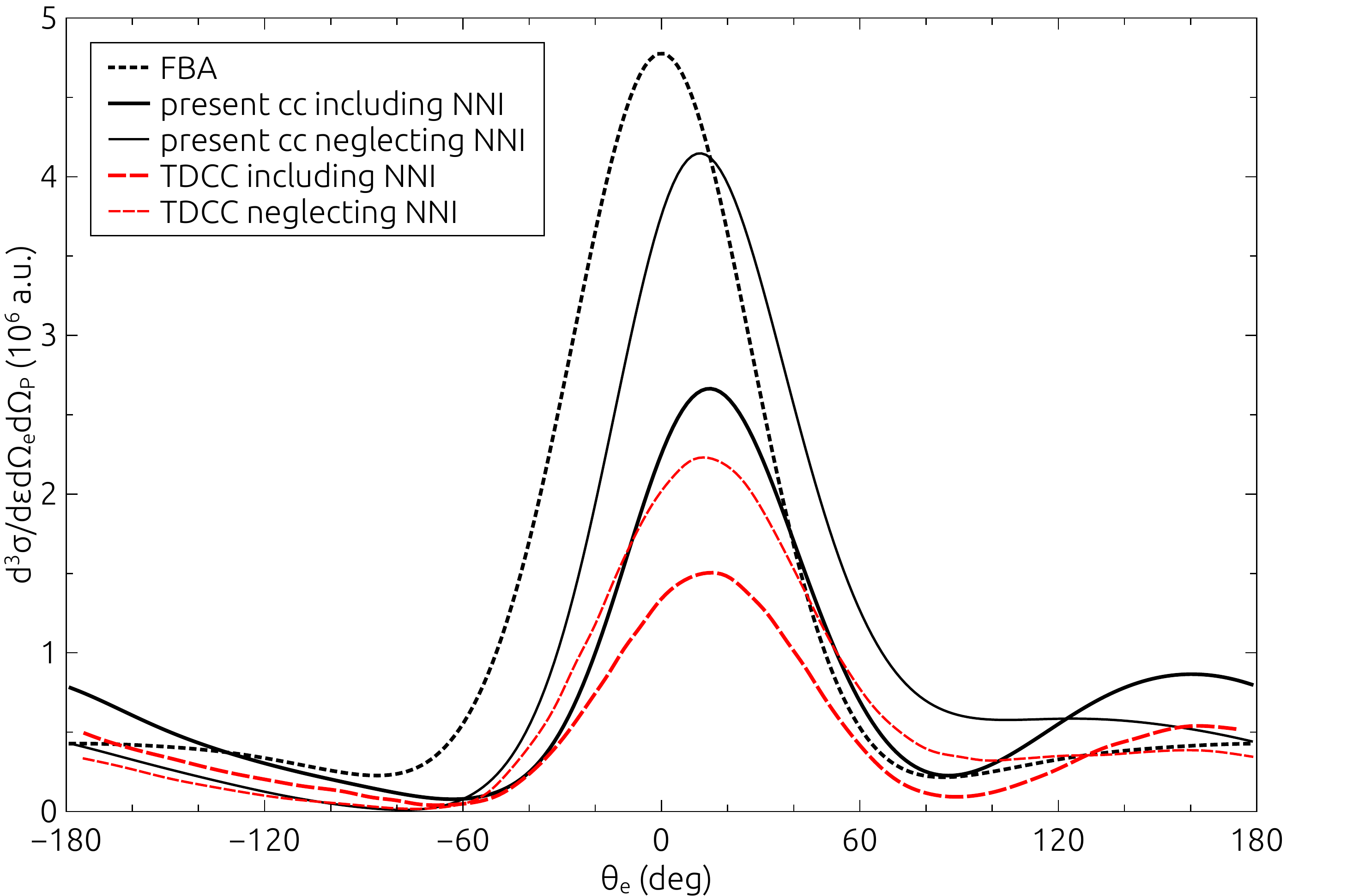}
\caption{The same as Fig.~\ref{fig:G1_200keV_04eV_02mrad}, but the results of calculation neglecting the NN interaction are also shown.}
\label{fig:G2_200keV_04eV_02mrad}
\end{figure}

\begin{figure}[htb] 
\includegraphics[width=0.98\linewidth]{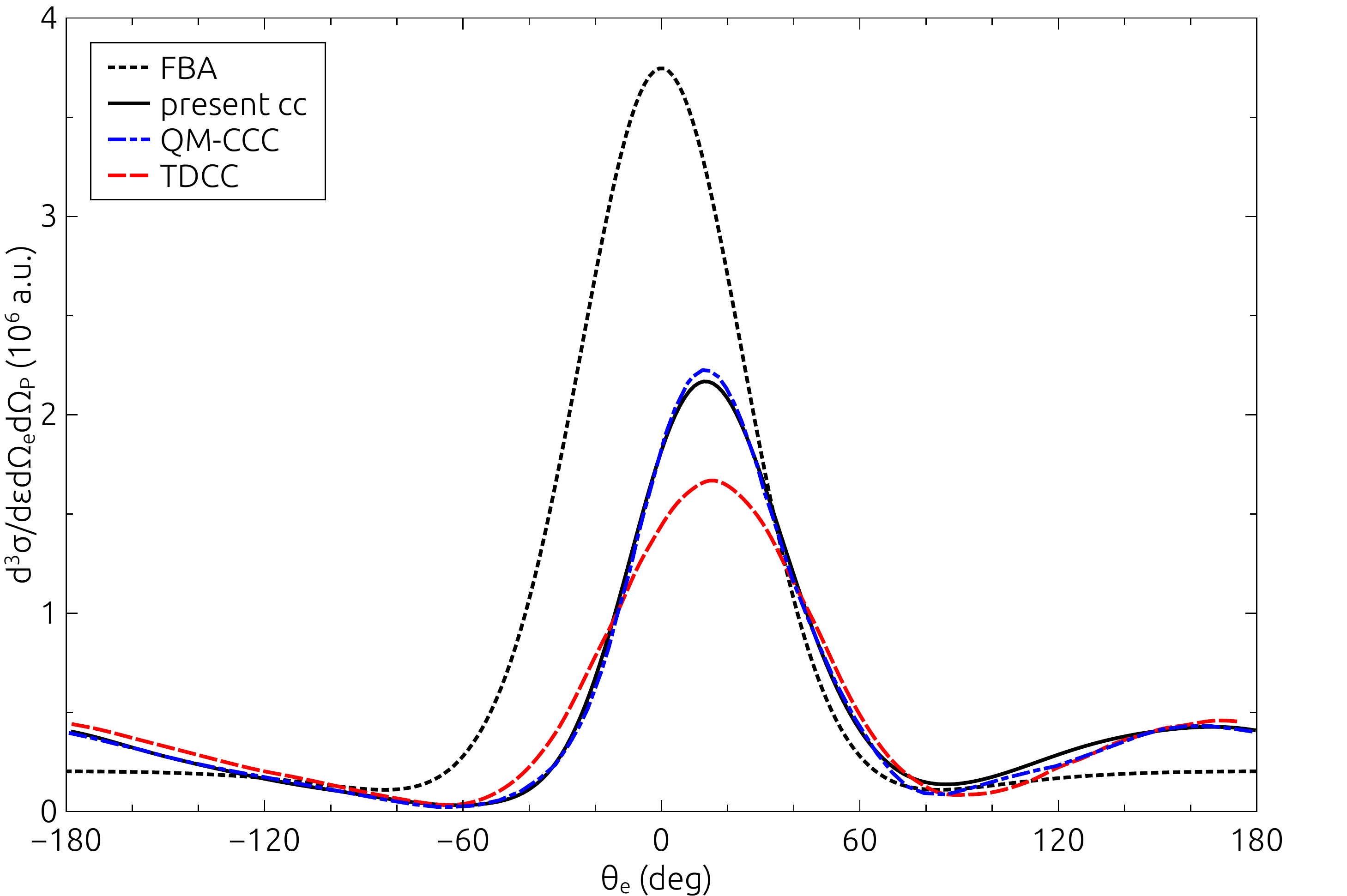}
\caption{TDCS for antiproton-impact ionization of hydrogen at $200$~keV in the scattering plane. The scattering angle of the projectile is $0.2$~mrad and the ejected electron energy is $7$~eV. The results of the QM-CCC and TDCC approaches are from Refs.~\cite{abd_11} and~\cite{cia_13}, respectively.}
\label{fig:G_200keV_02mrad_07eV}
\end{figure}

\begin{figure}[htb] 
\includegraphics[width=0.98\linewidth]{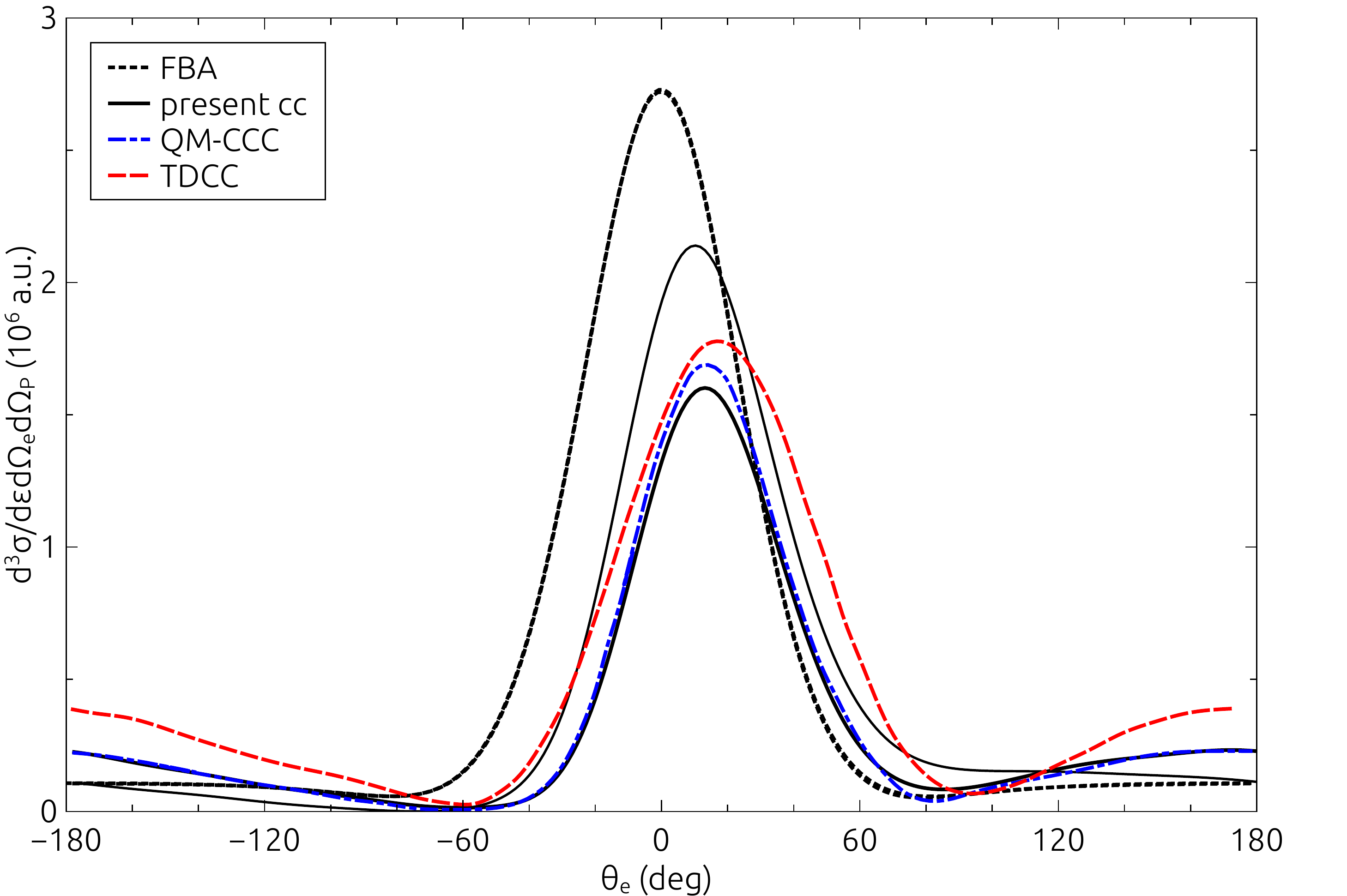}
\caption{TDCS for antiproton-impact ionization of hydrogen at $200$~keV in the scattering plane. The scattering angle of the projectile is $0.2$~mrad and the ejected electron energy is $10$~eV. The results of the QM-CCC and TDCC approaches are from Refs.~\cite{abd_11} and~\cite{cia_13}, respectively.}
\label{fig:G_200keV_02mrad_10eV}
\end{figure}

\begin{figure}[htb] 
\includegraphics[width=0.98\linewidth]{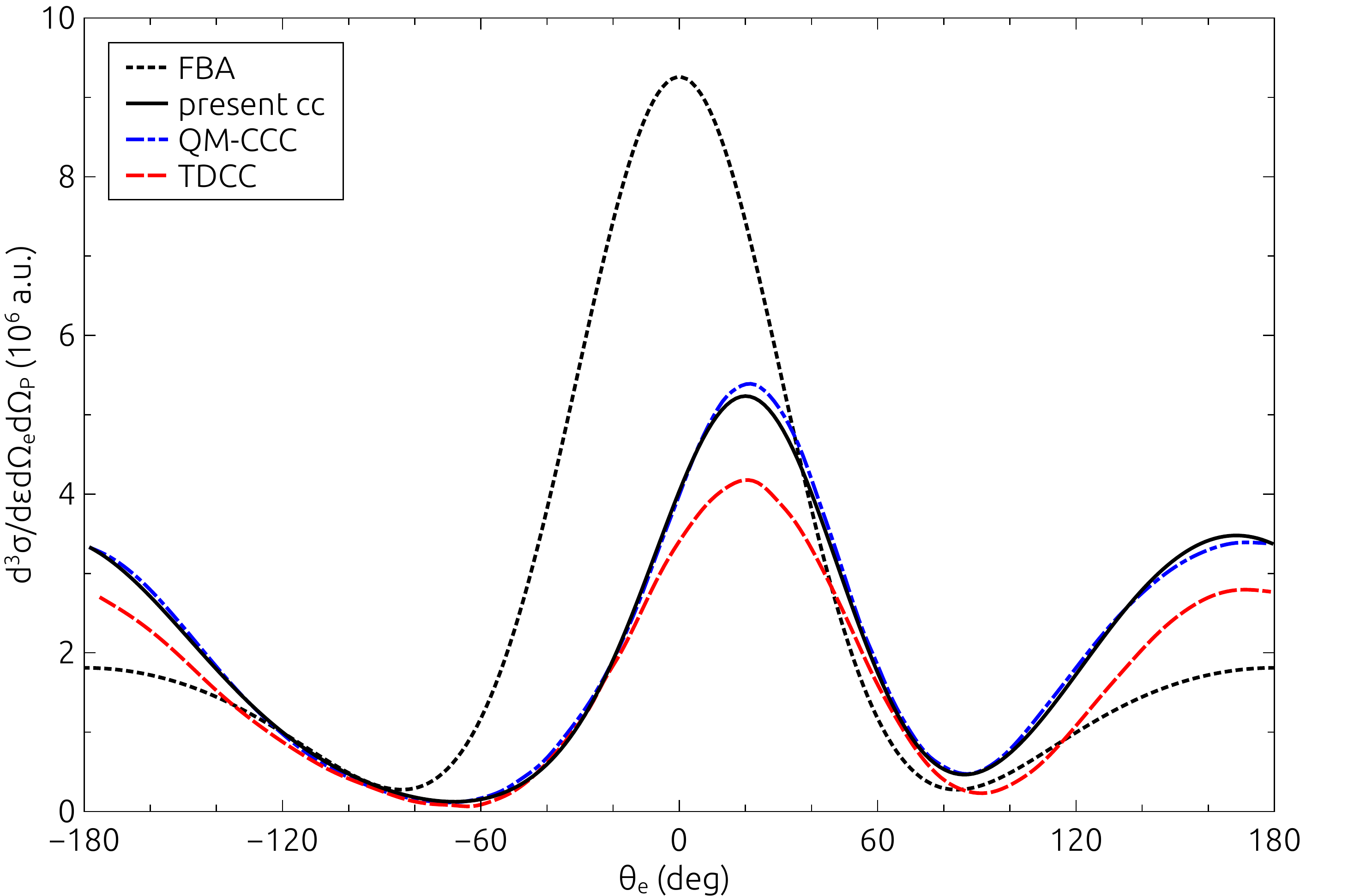}
\caption{TDCS for antiproton-impact ionization of hydrogen at $200$~keV in the scattering plane. The scattering angle of the projectile is $0.1$~mrad and the ejected electron energy is $4$~eV. The results of the QM-CCC and TDCC approaches are from Refs.~\cite{abd_11} and~\cite{cia_13}, respectively.}
\label{fig:G_200keV_04eV_01mrad}
\end{figure}

\begin{figure}[htb] 
\includegraphics[width=0.98\linewidth]{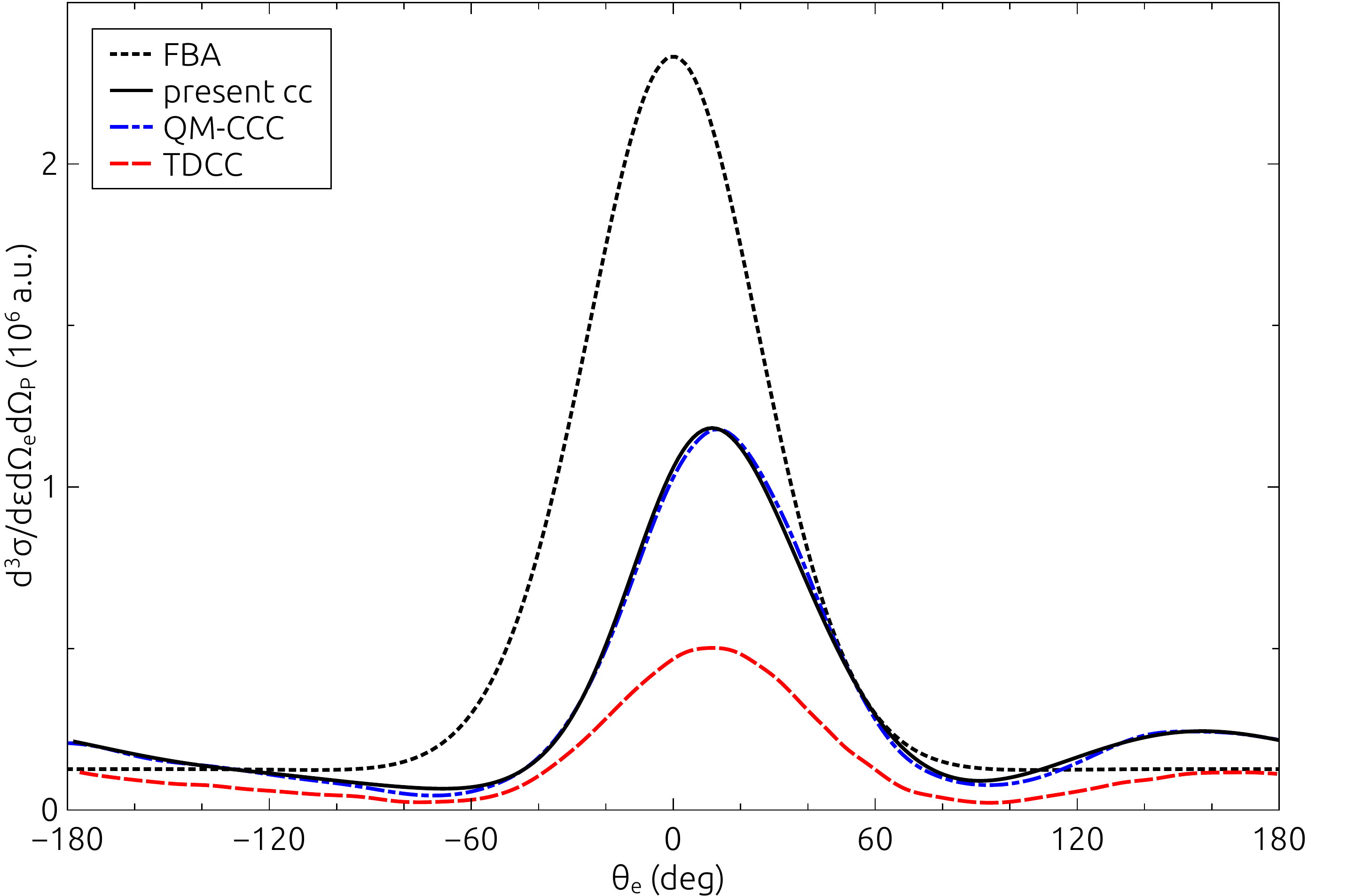}
\caption{TDCS for antiproton-impact ionization of hydrogen at $200$~keV in the scattering plane. The scattering angle of the projectile is $0.3$~mrad and the ejected electron energy is $4$~eV. The results of the QM-CCC and TDCC approaches are from Refs.~\cite{abd_11} and~\cite{cia_13}, respectively.}
\label{fig:G_200keV_04eV_03mrad}
\end{figure}

\begin{figure}[htb] 
\includegraphics[width=0.98\linewidth]{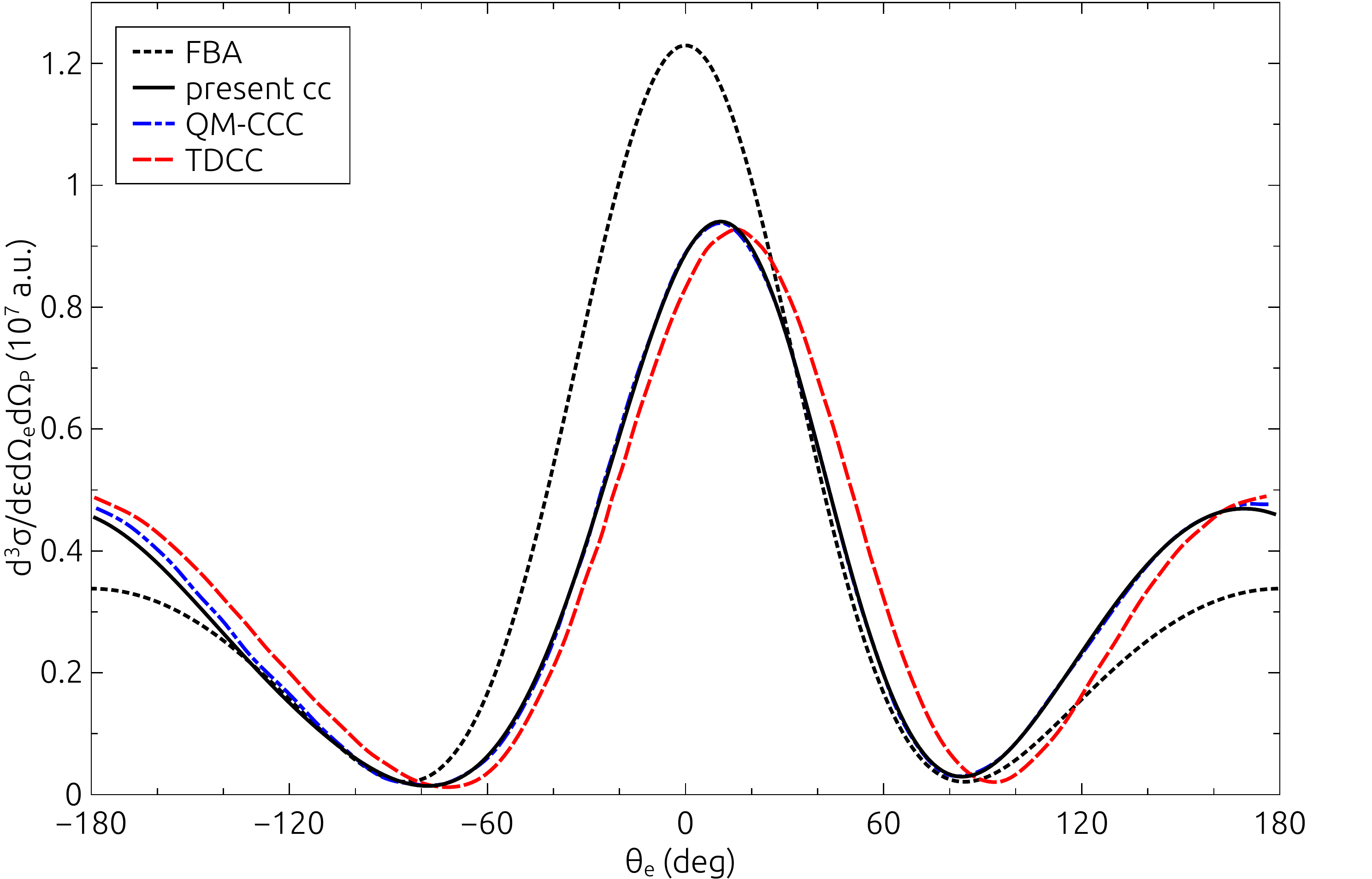}
\caption{TDCS for antiproton-impact ionization of hydrogen at $500$~keV in the scattering plane. The momentum transfer $q$ is $0.25$~a.u. and the ejected electron energy is $5$~eV. The results of the QM-CCC and TDCC approaches are from Refs.~\cite{abd_11} and~\cite{cia_13}, respectively.}
\label{fig:G_500keV_05eV_020au}
\end{figure}

\begin{figure}[htb] 
\includegraphics[width=0.98\linewidth]{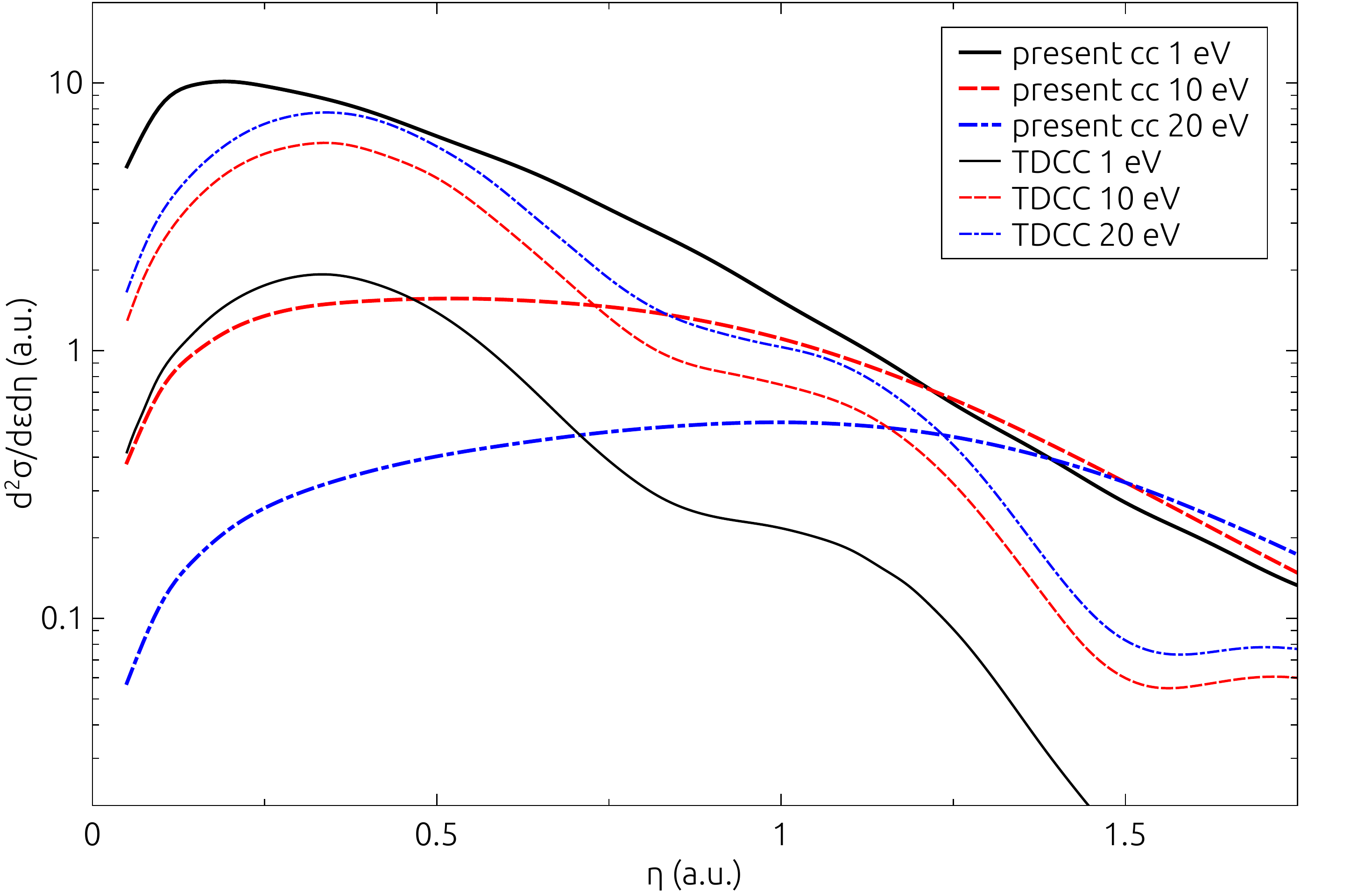}
\caption{DDCS as a function of the transverse component of momentum transfer for an energy of the ejected electron of $1$, $10$, and $20$~eV at an antiproton incident energy of $200$~keV. The results of the TDCC approach are from Refs.~\cite{cia_13}.}
\label{fig:G_DDCS_q_200keV}
\end{figure}

\begin{figure}[htb] 
\includegraphics[width=0.98\linewidth]{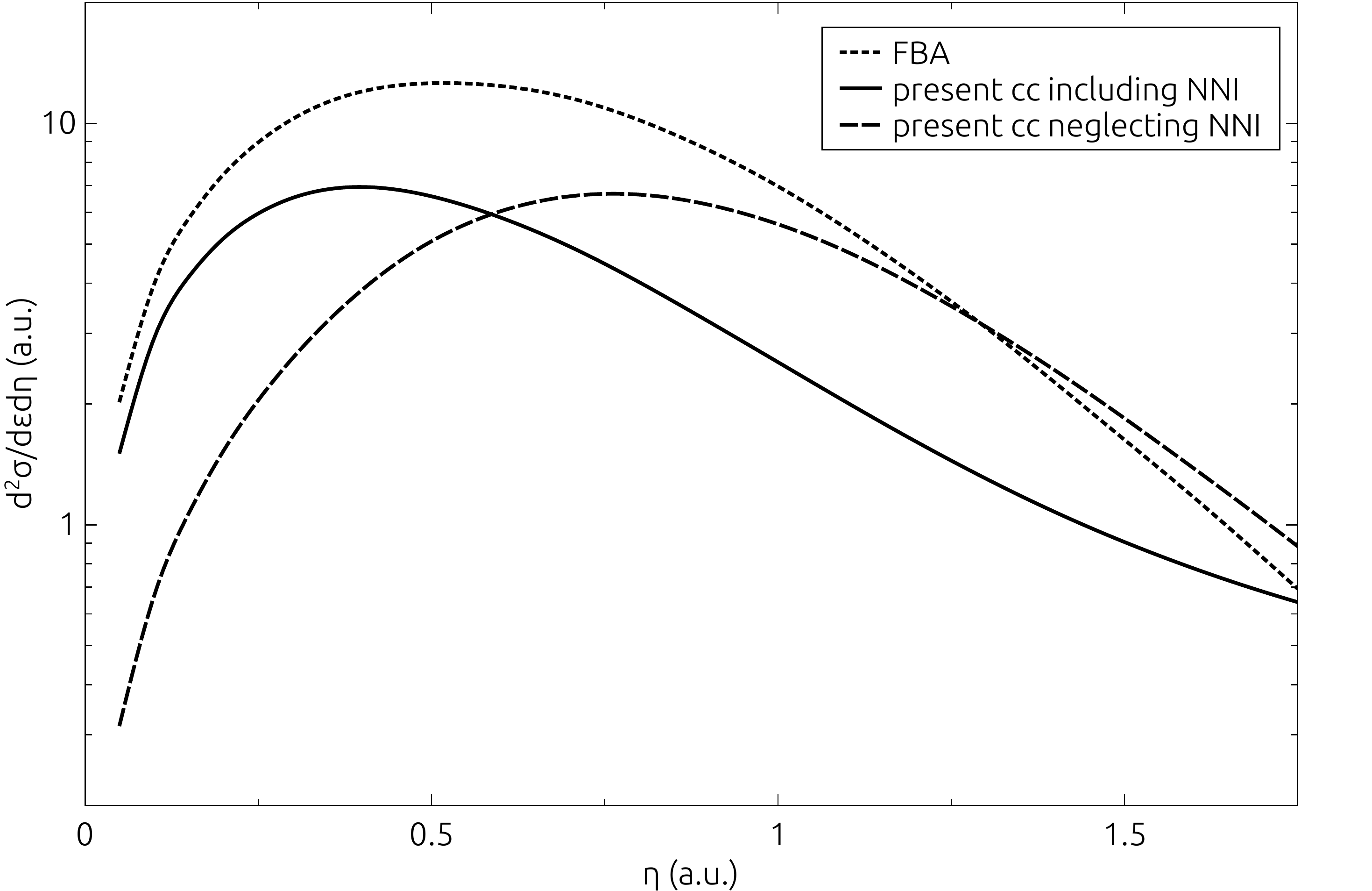}
\caption{DDCS as a function of the transverse component of momentum transfer for an ejected electron of $5$~eV at an antiproton incident energy of $30$~keV.}
\label{fig:G_DDCS_q_30keV}
\end{figure}

\begin{figure}[htb] 
\includegraphics[width=0.98\linewidth]{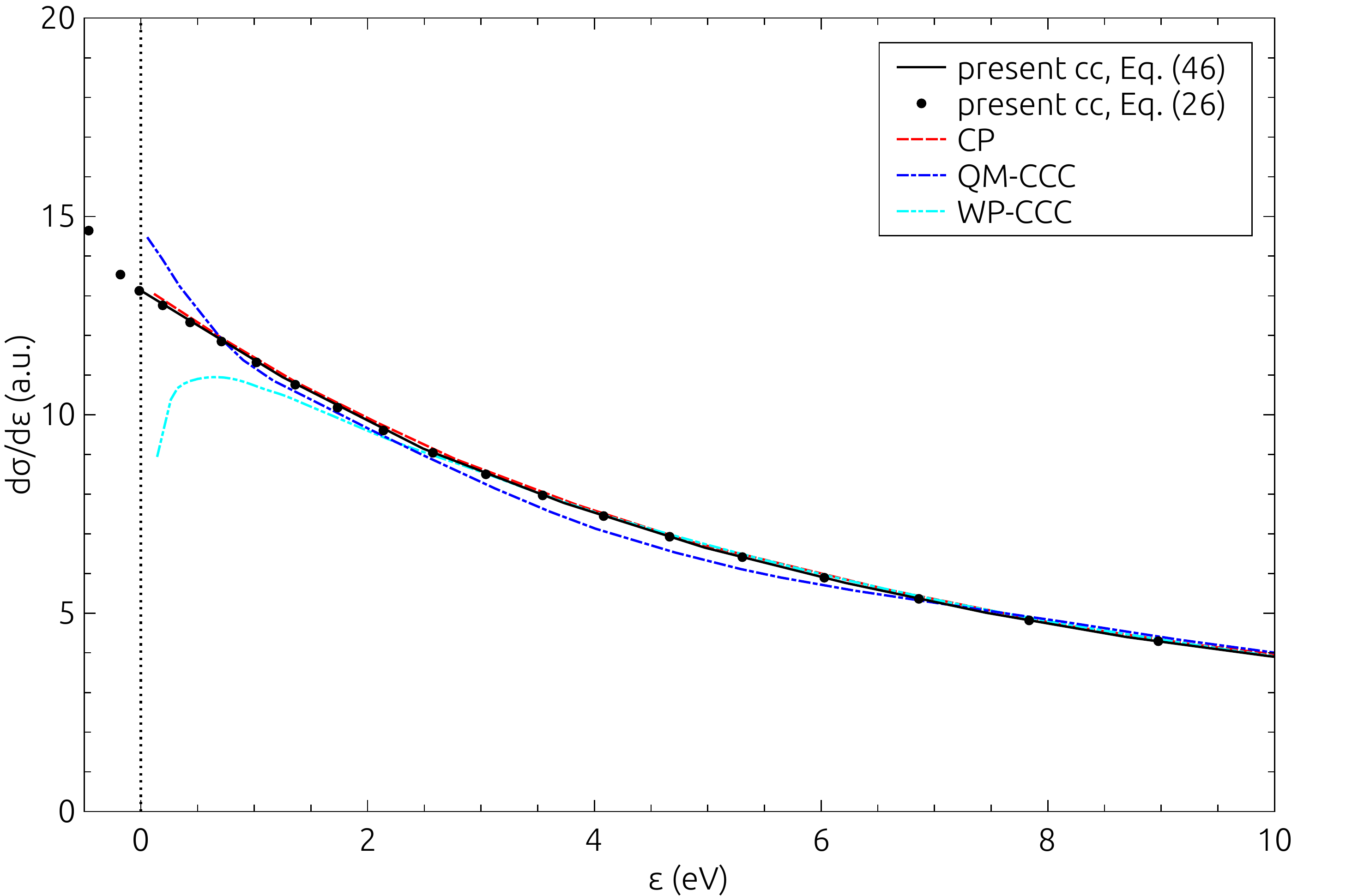}
\caption{SDCS at an incident antiproton energy of 30~keV. The results of the CP, QM-CCC and WP-CCC approaches are from Refs.~\cite{mcg_10}, \cite{abd_11}, and~\cite{abd_16}, respectively.}
\label{fig:G_SDCS_30keV}
\end{figure}

Let us start with presenting the total ionization cross sections. In Table~\ref{tab:TICS}, the present results of the full coupled-channel (cc) as well as corresponding FBA mode calculations, obtained by Eq.~\eqref{eq:total_p_ion} are compared with the results of the non-perturbative approaches of Refs.~\cite{abd_11a,cia_13} and the analytical FBA results~(see, e.g., Refs.~\cite{mcd_70,sar_14}).
\begin{table}[htb]
\caption{Total ionization cross sections (in units of $10^{-16}$~cm$^{-2}$) of atomic hydrogen under antiproton impact at various impact energies. The CCC and TDCC results are from Refs.~\cite{abd_11a} and~\cite{cia_13}, respectively.}
\label{tab:TICS}
\begin{center}
\begin{tabular}{|c|cc|ccc|} \hline \hline
Energy (keV) & Analytical FBA & Present FBA & CCC & TDCC & Present full  \\ \hline 
$30 $        & $2.15$         & $2.16$      & $1.35$		  & $1.46$	       & $1.37$        \\
$200$        & $0.77$         & $0.77$      & $0.66$		  & $0.65$             & $0.68$        \\ 
$500$        & $0.36$         & $0.36$      & $0.34$		  & $0.33$             & $0.35$        \\ \hline \hline 
\end{tabular}
\end{center}
\end{table}
Comparing the second and third columns of the table, one can see that the present FBA mode results are in excellent agreement with the analytical ones at all antiproton impact energies under consideration. The results of the full calculation are also in good agreement with the previous studies of Refs.~\cite{abd_11a,cia_13}. However, the results of Ciappina {\it et al.}~\cite{cia_13} at $30$~keV impact are noticeably larger than the present ones and the results of Ref.~\cite{abd_11a}.
\\
Briefly discussed the total ionization cross sections, we turn to the triply differential cross sections. Following Abdurakhmanov {\it et al.}~\cite{abd_11} and Ciappina {\it et al.}~\cite{cia_13}, we adopt their conventions. So the direction of the scattered projectile is fixed by the value of the momentum transfer $q$ or by the projectile deviation angle $\theta_{\rm P}$. The polar angle $\theta_e$ of the ejected electron runs from $-180\degree$ to $180\degree$ relative to the direction of the momentum transfer. The electron emission is considered in the scattering plane only.
\\
Fig.~\ref{fig:G1_200keV_04eV_02mrad} shows the TDCS for ionization of atomic hydrogen by impact of $200$-keV antiprotons with a scattering angle of $0.2$~mrad and for an ejected electron energy of $4$~eV.
The results of the non-perturbative QM-CCC~\cite{abd_11} and TDCC~\cite{cia_13} approaches along with the FBA analytical results and the present FBA mode results (see Eq.~\eqref{eq:B1_coef}) are also shown. In the figure, one can see perfect agreement between the FBA mode results and the analytical FBA results, which in the following will be labeled as FBA without indicating the type. This agreement verifies the convergence of our results in the FBA mode as well as in the full calculation. All displayed curves demonstrate the two-peak structure with the binary peak in the direction of the momentum transfer and the recoil peak in the opposite direction. Note that the FBA TDCS are always symmetric with respect to the momentum transfer direction. Comparing to the FBA, all presented non-perturbative theories predict the reduced binary and enhanced recoil peaks both rotated away from the direction of the scattered antiproton. For both peaks the expected positions agree with each other, however, there is a noticeable discrepancy in the magnitude. The present results being in good agreement with the QM-CCC results lie significantly above the TDCC results. Ciappina {\it et al.}~\cite{cia_13} assumed that it is the non-perturbative treatment of the higher-order electron-projectile terms of close-coupling formalisms rather than the NN interaction effect, as it was proposed by Abdurakhmanov {\it et al.}~\cite{abd_11}, which leads to the shift of the binary and recoil peaks relatively to the FBA results. In our semiclassical calculations the NN interaction is treated as the phase factor in Eq.~\eqref{eq:coef_four_trans}, i.e. in the same way as in the TDCC calculations of Ref.~\cite{cia_13}. Thus we are also able to examine the role of the NN interaction by taking it into account or ignoring in the performed calculations.
\\
In Fig.~\ref{fig:G2_200keV_04eV_02mrad}, we display the TDCS for the same parameters as in Fig.~\ref{fig:G1_200keV_04eV_02mrad}, together with the results of the calculation neglecting the NN interaction ($\delta(b) \equiv 0$ in Eq.~\eqref{eq:NN_phase}). 
The corresponding results of Ref.~\cite{cia_13} and the FBA results are also shown. Here we indeed see that inclusion of the NN interaction does not affect the position of the binary peak, in accordance with the suggestion of Ref.~\cite{cia_13}. Moreover, in our calculation the inclusion of the NN interaction also significantly reduces the TDCS. However, the peak value of the present TDCS obtained in the calculation ignoring the NN interaction is about $10\%$ smaller than the FBA result, whereas the peak value of the TDCC TDCS~\cite{cia_13} is only about $50\%$ of the FBA result.
\\
In Figs.~\ref{fig:G_200keV_02mrad_07eV} and~\ref{fig:G_200keV_02mrad_10eV}, the TDCS for higher electron ejection energies of $7$ and $10$~eV, respectively, are presented. 
For every curve in these figures, the overall form is the same as for a lower energy ejection of $4$~eV (see Fig.~\ref{fig:G1_200keV_04eV_02mrad}) and the positions of the binary and recoil peaks are nearly unchanged. One again can see good agreement between the present and QM-CCC results of Ref.~\cite{abd_11}, which are almost indistinguishable except for the binary-peak maximum at about $13\degree$ and the minimum at about $86\degree$. The small differences at these regions increase with increasing the energy of the ejected electron. The binary peak positions of the TDCC TDCS of Ref.~\cite{cia_13} agree with the present for both energies, however, there is again the inconsistency in the magnitude. Moreover, the TDCC TDCS increase with increasing the energy of the ejected electron in contradiction with the other theories.
\\
Next, following Refs.~\cite{abd_11,cia_13}, we investigate the TDCS for various projectile scattering angles. The results are shown in Figs.~\ref{fig:G_200keV_04eV_01mrad} and~\ref{fig:G_200keV_04eV_03mrad}.
For all presented non-perturbative theories the magnitude of the binary peak decreases with increasing the projectile scattering angle, in accordance with the FBA. The position of the binary peak shifts to its FBA position with increasing the projectile scattering angle. Here we again see the mismatch between the present and QM-CCC results from the one hand and the TDCC results from the other hand. This mismatch grows with increasing the projectile scattering angle.
\\
The TDCS at an antiproton incident energy of $500$~keV are shown in Fig.~\ref{fig:G_500keV_05eV_020au}. 
Note that at such a high impact energy, the FBA TDCS still differs from the non-perturbative ones, while the total ionization cross sections predicted by all approaches agree much better with each other (see Table~\ref{tab:TICS}). Here the results of the present approach, QM-CCC and TDCC agree in magnitude. However, in contrast to the previously discussed examples for the $200$~keV impact, the binary peak of TDCC TDCS is slightly shifted to the right compared to the present results and QM-CCC data. This may be caused by the inconsistency in the main text and the caption of Fig.~3 in Ref.~\cite{cia_13}. In the caption, it is stated that TDCS is plotted for the value of the total momentum transfer $q = 0.25$~a.u., while in the main text, that for the value of the transverse component of the momentum transfer $q_\perp \equiv \eta = 0.25$~a.u., which corresponds to the antiproton scattering angle $\theta_{\rm P} = 0.061$~mrad indicated there. The angle $\theta_f$ between the direction of the final projectile momentum $\vec{k_f}$ and the direction of the momentum transfer $\vec{q}$ equals to $52.3\degree$ and $58.6\degree$ for $q = 0.25$~a.u. and $\eta = 0.25$~a.u., respectively. We would like to point out that the TDCS for this kinematical regime has been first calculated within the CP method by McGovern {\it et al.}~\cite{mcg_09}. However, it is almost indistinguishable from the QM-CCC results of Ref.~\cite{abd_11}.
\\
The DDCS in energy of the ejected electron and transverse component of the projectile momentum transfer $\frac{d^2\sigma}{d\eps \, d\eta}$  for various energies of the ejected electron as a function of the transverse component of the projectile momentum transfer~$\eta$ at an incident antiproton energy of $200$~keV is presented in Fig.~\ref{fig:G_DDCS_q_200keV}. The results of the TDCC approach of Ref.~\cite{cia_13} are also shown.
It is clearly seen that for all energies of the ejected electron the present results disagree with the TDCC results both in the magnitude and shape. For small values of $\eta$, the present results for low energies of the ejected electron are larger than those for high ejection energies. For large values of $\eta$, the picture is inverted in accordance with the FBA, which is not shown here. It means that for large values of the momentum transfer the maximum of the DDCS is shifted from zero emission energy. For example, for $\eta = 1.75$~a.u. this maximum is located about $\eps = 25$~eV in the FBA. In contrast, the TDCC results for high energy of the ejected electron are larger than those for low energy in the whole range of the momentum transfer. The DDCS $\frac{d^2\sigma}{d\eps \, d\eta}$ being integrated over $\eta$ gives the SDCS $\frac{d\sigma}{d\eps}$, which in this case unexpectedly increases with increasing the energy of the ejected electron. The TDCC DDCS indicate also pronounced structures in the variation of $\eta$, which are not observed in our results.
\\
At a higher antiproton incident energy of $500$~keV, Ciappina~{\it et al.}~\cite{cia_13} found similar patterns as shown in Fig.~\ref{fig:G_DDCS_q_200keV} for the $200$~keV impact. However, these patterns are still too far from ours, which are very close to the FBA results and are not shown here.
\\
In order to explore the role of the NN interaction, it is more useful to consider DDCS at lower projectile incident  energies. In Fig.~\ref{fig:G_DDCS_q_30keV}, we display the DDCS $\frac{d^2\sigma}{d\eps \, d\eta}$ as a function of transverse momentum transfer $\eta$ for an ejected electron energy of $5$~eV at an antiproton incident energy of $30$~keV.
It is obvious that the DDCS is strongly influenced by the NN interaction in the whole range of the momentum transfer. However, no oscillatory structures in the variation of $\eta$ are observed again. 
The reason of the strong contradiction between the present and TDCC results for DDCS is unclear to us.
\\
Fig.~\ref{fig:G_SDCS_30keV} shows the SDCS in energy of the ejected electron $\frac{d\sigma}{d\eps}$ at an incident antiproton energy of $30$~keV together with the results of the non-perturbative approaches of Refs.~\cite{mcg_10,abd_11,abd_16}. 
The results of all approaches are in good agreement for the electron ejection energies larger than $7$~eV. However, the low-energy behavior is different. The WP-CCC SDCS of Abdurakhmanov~{\it et al.}~\cite{abd_16} has a maximum away from the zero emission energy, contrary to the other results. The present SDCS calculated using Eq.~\eqref{eq:dsig_de} is in excellent agreement with the CP results of McGovern~{\it et al.}~\cite{mcg_09}, and monotonically increases with decreasing the electron ejected energy. In order to verify this behavior, we also calculated SDCS using Eq.~\eqref{eq:dsig_de_discr}, which is also valid for negative energies, where it describes the excitation rather than the ionization process. For positive energies, the results obtained by Eq.~\eqref{eq:dsig_de_discr} are in perfect agreement with the results obtained by Eq.~\eqref{eq:dsig_de}, and smoothly increase with decreasing the energy below the ionization threshold. This smooth transition between the excitation to high-energy bound states and the ionization to low-energy continuum states is quite reasonable from a general point of view.\\
The WP-CCC method recently developed by Abdurakhmanov~{\it et al.}~\cite{abd_16} is formulated in the framework of the single-center semiclassical convergent close coupling approach. The key feature of the method is using stationary wave packets for discretization of the continuous spectrum of the target. Such continuum discretization allows one to generate pseudostates with arbitrary energies and distribution.
The reason of the low-energy fall of the WP-CCC results might be a poor implementation of the wave packets describing low-energy states. By construction, the wave packets form an orthonormal basis for positive-energy states. However, a low-energy wave packet of a fine width has a huge size in the coordinate space. This requires the upper limit of the integration over the radial variable in the calculation of the matrix elements to be very large, which is hard to achieve. Furthermore, the results of the FBA mode calculation are determined by the matrix element involving the initial rather localized ground state, and thus are insensitive to the shape of the final state wave packet at large distances. This might be a reason of the good agreement of the WP-CCC results in the FBA mode with the analytical FBA predictions (see Fig.~9 in Ref.~\cite{abd_16}).
\subsection{C$^{6+}$-impact ionization of hydrogenlike xenon ion} \label{ss:C-Xe}
\begin{figure}[htb] 
\includegraphics[width=0.98\linewidth]{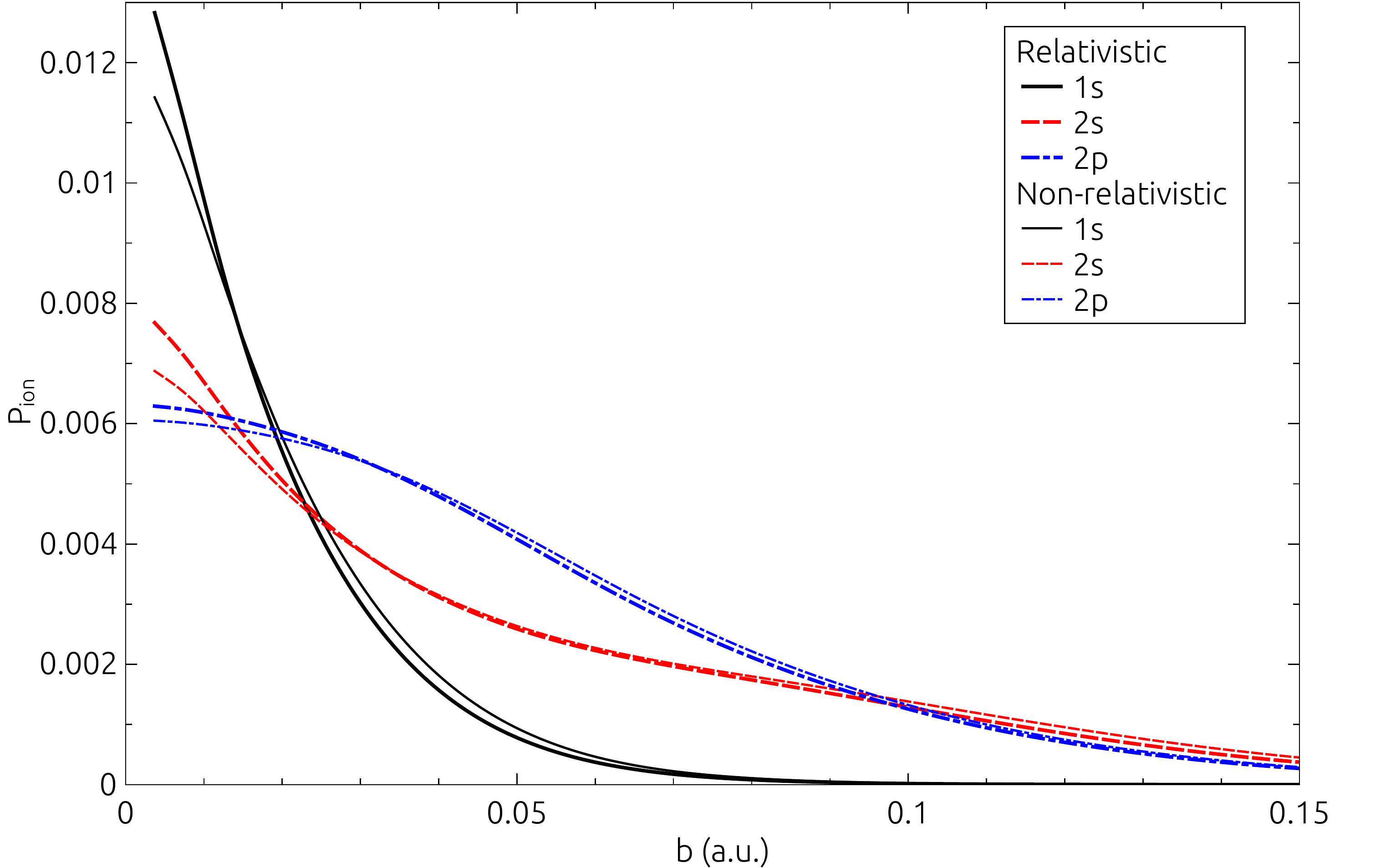}
\caption{Impact-parameter dependence of the total ionization probability in the $100$~MeV/u C$^{6+}$-Xe$^{53+}$ collision for various initial states. The results of the calculation in the non-relativistic limit are also shown.}
\label{fig:C-Xe_KL}
\end{figure}

\begin{figure}[htb] 
\includegraphics[width=0.98\linewidth]{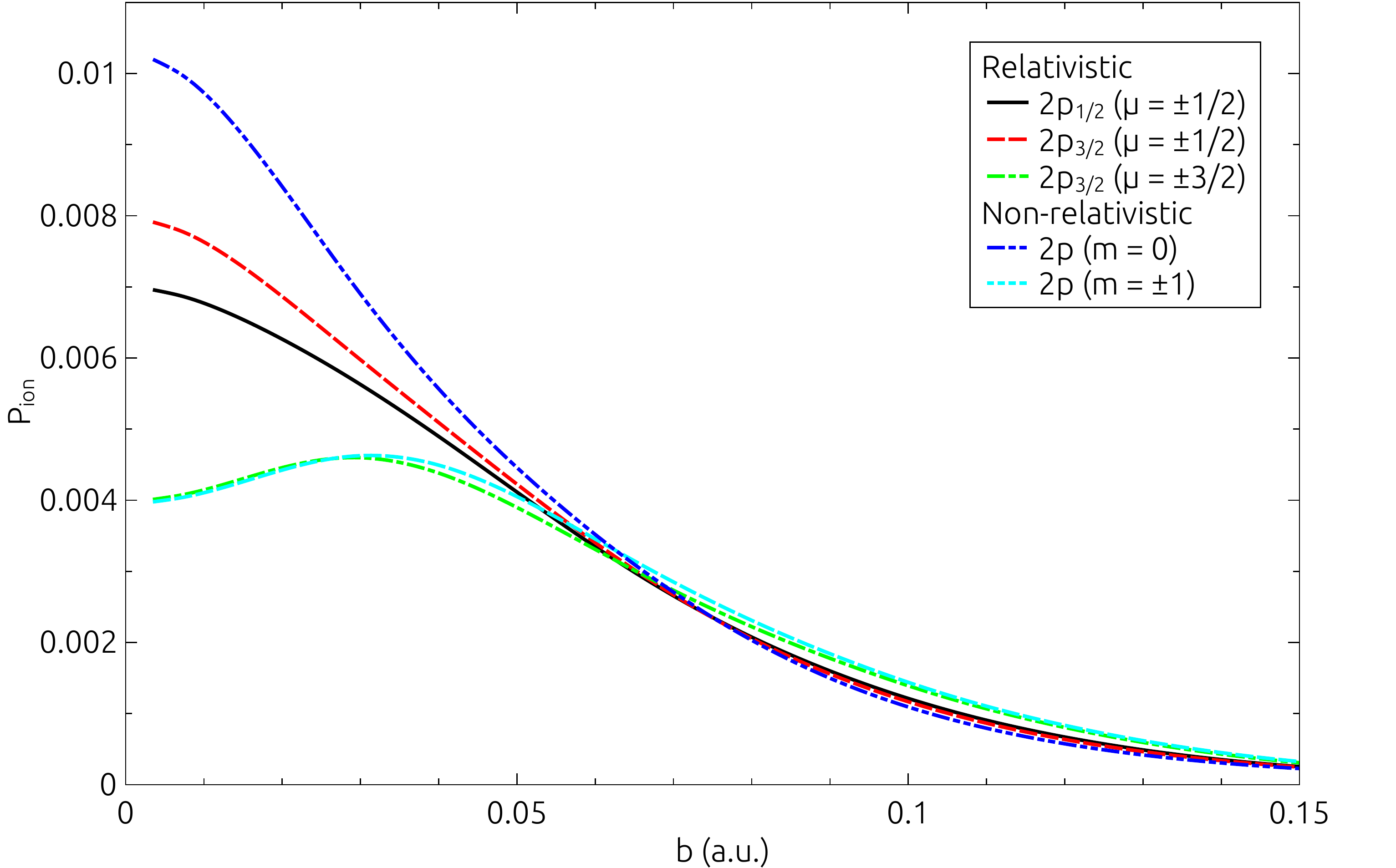}
\caption{Impact-parameter dependence of the total ionization probability in the $100$~MeV/u C$^{6+}$-Xe$^{53+}$ collision for various momentum projections of the initial $2p$ state. The results of the calculation in the non-relativistic limit are also shown.}
\label{fig:C-Xe_mu}
\end{figure}

Finally, in order to examine relativistic effects, we have considered the $100$~MeV/u C$^{6+}$-Xe$^{53+}$ collision. The $100$~MeV/u carbon nuclei have already been used to study the fully differential cross sections for single ionization of helium atom~\cite{sch_03}. The impact-parameter dependencies of the total ionization probabilities from the $K$- and $L$-shells have been calculated. In order to explore the relativistic effects induced by a large target charge, we also carried out the calculation in the non-relativistic limit,  where the standard value of the speed of light $c$ was multiplied by a factor of $1000$. The comparison of the results of both calculations is shown in Fig.~\ref{fig:C-Xe_KL}.
From the figure, one can see that the relativistic effects enhance the total ionization probability at small impact parameters and reduce it at large ones for all considered states. Also noticeable is the dominance of the ionization from the $1s$ state at small impact parameters. In contrast to the ionization from the $1s$ and $2s$ states, the total ionization probability from the $2p$ states, averaged over the values of total angular momentum and its projections, is convex upwards at small impact parameters.
\\
It is also worth to consider the impact-parameter dependence of the total ionization probability from the $2p$ states with various quantum numbers $j$ and $\mu$. These results together with the results of the corresponding non-relativistic calculation are shown in Fig.~\ref{fig:C-Xe_mu}.
The total ionization probability does not depend on the sign of the projections $\mu$ and $m$. However, for both calculations, it significantly depends on its absolute value. In the relativistic calculation, the total ionization probability depends also on the total angular momentum $j$ of the initial $2p$ state. 
\section{Conclusion}  \label{s:conclusion}
In this study, we have presented the relativistic semiclassical approach based on the Dirac equation to calculation of differential ionization cross sections in ion-atom collisions. {\it B}-splines are used to discretize the Dirac continua of the target. As the first test, the method has been applied to calculation of various differential cross sections for antiproton-impact ionization of atomic hydrogen. 
Several discrepancies in available results of non-perturbative approaches based on the Schr\"odinger equation have been resolved. We may assume that the TDCC calculations performed by Ciappina {\it et al.}~\cite{cia_13} have an issue at the stage of the Fourier transform from the $b$- to $\eta$-representation of the ionization amplitude. We also suppose that the low-energy behavior of the WP-CCC SDCS found by Abdurakhmanov {\it et al.}~\cite{abd_16} arises from the lack of normalization of thin wave packets with a small energy.
\\
The method has also been applied to explore the relativistic effects on the total ionization probability from the $K$- and $L$-shells of hydrogenlike xenon ion under the impact of carbon nuclei. The approach is also suitable for investigation of more complicated many-electron targets.
\\
In future, we plan to apply the developed approach to study ionization processes at the differential level in collisions involving heavy targets, where the relativistic effects are extremely important.
\section*{Acknowledgements}
We thank Alisher Kadyrov and Igor Bray for valuable discussions. This work was supported by RFBR (Grants No. 15-03-07644, No. 16-02-00233, and No. 17-52-53136), by SPSU (Grants No. 11.38.237.2015 and 11.65.41.2017), and by the President of the Russian Federation (Grant No. MK-6970.2015.2). A.I.B. acknowledges the support from the German-Russian Interdisciplinary Science Center (G-RISC) funded by the German Federal Foreign Office via the German Academic Exchange Service (DAAD) and the FAIR-Russia Research Center.
\bibliography{biblio}{}
\bibliographystyle{h-physrev}
\end{document}